\documentclass[mathleft,a4paper]{an}
\usepackage{graphicx}
\usepackage{times}
\usepackage{url}
\begin{document}

\def\yu{2005\,YU\ensuremath{_{55}}}
\def\bx{2012\,BX\ensuremath{_{34}}}

\Pagespan{1}{}
\Yearpublication{2012}%
\Yearsubmission{2012}%
\Month{11}%
\Volume{999}%
\Issue{88}%

\title{The Fly's Eye Camera System -- \\
        an instrument design for large \'etendue time-domain survey}

\author{%
Andr\'as P\'al\inst{1,2}\fnmsep\thanks{Corresponding author: \email{apal@flyseye.net}\newline}
\and
L\'aszl\'o M\'esz\'aros\inst{1,2}
\and
Gergely Cs\'ep\'any\inst{1}
\and
Attila Jask\'o\inst{1}
\and
Ferenc Schlaffer\inst{3}
\and  
Kriszti\'an Vida\inst{1}
\and
Gy\"orgy Mez\H{o}\inst{1}
\and
L\'aszl\'o D\"obrentei\inst{1}
\and 
Ern\H{o} Farkas\inst{1}
\and
Csaba Kiss\inst{1}
\and
Katalin Ol\'ah\inst{1}
\and
Zsolt Reg\'aly\inst{1}
}
\titlerunning{The Fly's Eye Camera System}
\authorrunning{A. P\'al et al.}
\institute{
MTA Research Centre for Astronomy and Earth Sciences, Konkoly Thege Mikl\'os \'ut 15-17, Budapest, H-1121, Hungary
\and 
Department of Astronomy, Lor\'and E\"otv\"os University, P\'azm\'any P\'eter s\'et\'any 1/A, Budapest H-1117, Hungary
\and
MTA Research Centre for Astronomy and Earth Sciences, Csatkai E. u. 6-8, Sopron, H-9400, Hungary}

\received{15 Nov 2012}
\accepted{15 Nov 2012}
\publonline{later}


\abstract{%
In this paper we briefly summarize the design concepts of the 
{\it Fly's Eye Camera System}, 
a proposed high resolution all-sky monitoring
device which intends to perform high cadence time domain astronomy
in multiple optical passbands while
still accomplish a high \'etendue. Fundings have already been accepted
by the Hungarian Academy of Sciences
in order to design and build a {\it Fly's Eye} device unit. Beyond the
technical details and the actual scientific goals, this paper
also discusses the possibilities and yields of a network operation
involving $\sim10$ sites distributed geographically in 
a nearly homogeneous manner. Currently, we expect 
to finalize the mount assembly -- that performs the sidereal tracking
during the exposures -- until the end of 2012 and to 
have a working prototype with a reduced number of individual cameras
sometimes in the spring or summer of 2013.}

\maketitle

\section{Introduction}
\label{sec:introduction}

Astrophysical phenomena take place on a wide range of timescales.
From the shortest millisecond signals of pulsars up to the lifetime of
stars, that can be comparable to the age of the Universe, there is an 
astonishing span of $\sim20$ magnitudes. The key to unveil the 
physical processes beyond these phenomena is to monitor the 
alterations of observable quantities, such as flux.
Although some of the processes have 
their own characteristic timescales, most of the complex systems 
exhibit variations on a broader temporal spectrum. These complex
systems show signs of periodic, quasi-periodic and
sudden transient, eruptive processes. The observed timescales 
imply not only the possible durations of matter rearrangement whatever 
is the reason behind, but constrain 
the physical backgrounds of the variabilities of the observed 
systems. Hence, 
persistent monitoring of such ``astrophysical laboratories'' helps us
to understand how stars evolve, and from a wider perspective, 
how planetary systems and even our Solar System develop from their early 
stages of life until its end. 

Astronomical surveys require a complex optical and detector system 
to cover a large field-of-view (FOV), which often pairs with large 
light collecting area. The cumulative light collecting power, 
known as the \emph{\'etendue} defines how effective
a certain instrument is for survey purposes. 
By following the astronomical scientific 
discovery orientations, as one of these is the time-domain
astronomy (see \cite{blandford2010}), recent initiatives for survey projects
highly focus on the most extensive ways of implementing 
instrumentation with high optical acceptance.

The aim of our plan is to develop and build an instrument 
coined as {\it Fly's Eye Camera System} that allows 
the continuous monitoring of optical sky variability. The timescale window
in which the instrument will operate covers $\sim 6$ 
order of magnitudes: from the data acquisition cadence 
in the range of  minutes up to the expected range of several years of operation. 

The proposed design yields an \'etendue 
that is comparable to the currently operating survey programs,
such as the highly successful \emph{Kepler} space telescope (\cite{borucki2007})
and the ambitious \emph{Pan-STARRS} project (\cite{kaiser2002}). 
The \emph{Fly's Eye Camera System} is a 
``high cadence + low imaging resolution + large solid angle coverage'' 
instrument. Unlikely to e.g. \emph{Kepler} that uses a 
``high cadence + high imaging resolution + small solid angle coverage'' setup and
\emph{Pan-STARRS} that provides a
``small cadence + high imaging resolution + large solid angle coverage'' combination, 
the {\it Fly's Eye} allows the monitoring of a presently unexplored range of 
the domain of astronomical events.

Persistent monitoring of several thousand bright, scientifically 
relevant systems can only be implemented by the means of smaller 
multiplexed instruments exploiting smaller imaging resolutions.
In addition, transient detection in various known or undiscovered systems and
statistical analysis is also feasible in this domain.
Thus, such an instrument will provide a backbone of high resolution 
photometric, polarimetric, interferometric, infrared, spectroscopic 
and space-borne follow-up measurements as well. 
An extended, nearly uniform geographical distribution 
of $8-10$ {\it Fly's Eye} units would result in a 
light-grasp power comparable to the \emph{Large Synoptic Survey Telescope} 
(LSST, \cite{ivezic2008}). The \emph{LSST} is the 
highest ranked ground based facility in the strategic 
roadmap of American astronomy for the next decade (\cite{blandford2010}).
Furthermore, as it is discussed in more details later on, the 
{\it Fly's Eye} design will provide a continuous transition to
the brighter targets from the fainter ones aimed to be observed by \emph{LSST}.

As we will explain here, the design is both simple and robust 
(Sec.~\ref{sec:instrumentdesign}) to build a geographically 
extended  network of this camera system, providing a more dense phase coverage
of the observed events and a wider perspective to the 
sky~(Sec.~\ref{sec:furtherdevelopment}).
The proposed scientific applications cover disciplines from 
within the nearby Solar System (including even the atmosphere 
of the Earth) up to extragalactic investigations (Sec.~\ref{sec:science}). 

\begin{figure*}
\begin{center}
\resizebox{!}{60mm}{\includegraphics{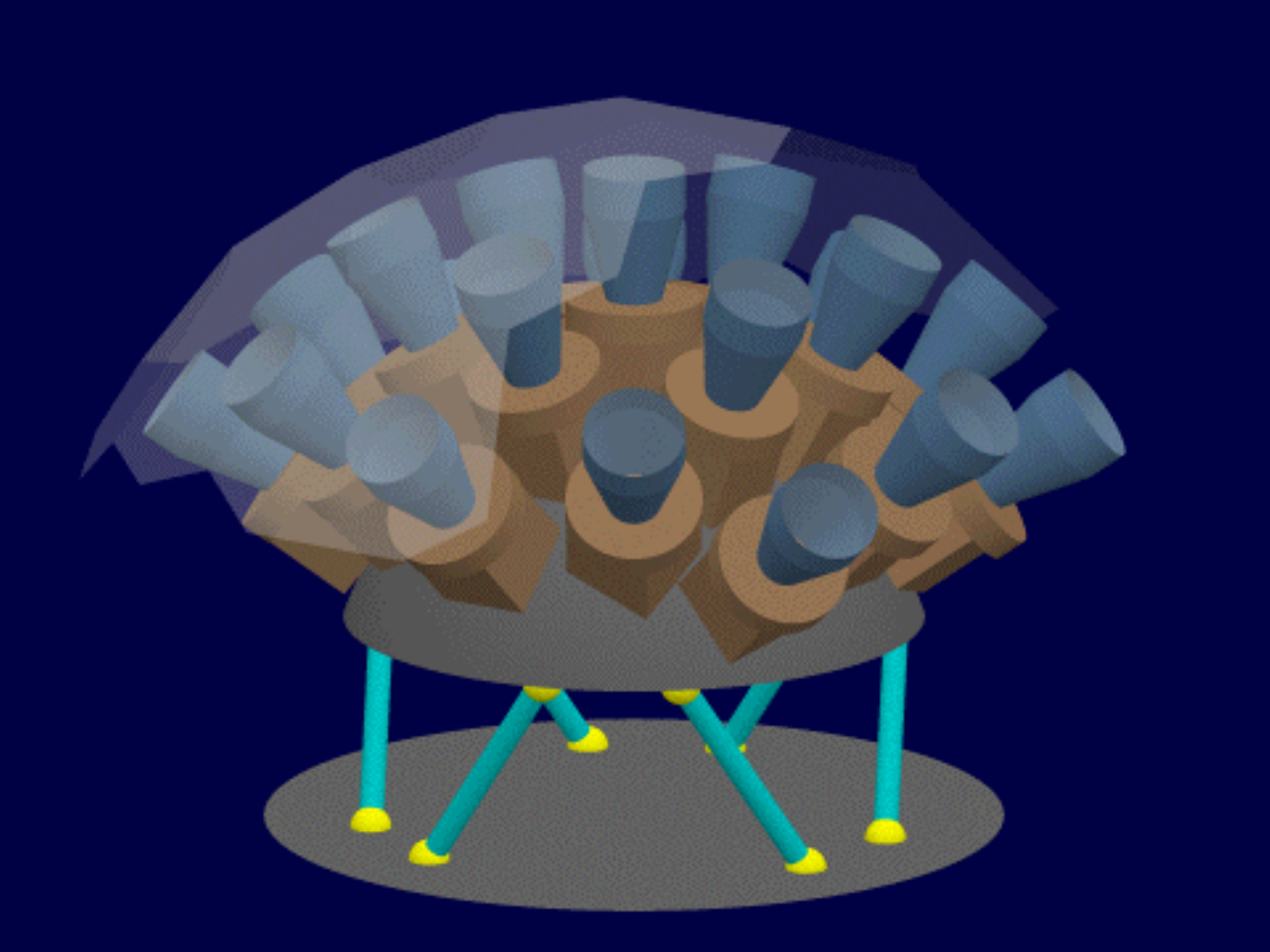}}\hspace*{4mm}
\resizebox{!}{60mm}{\includegraphics{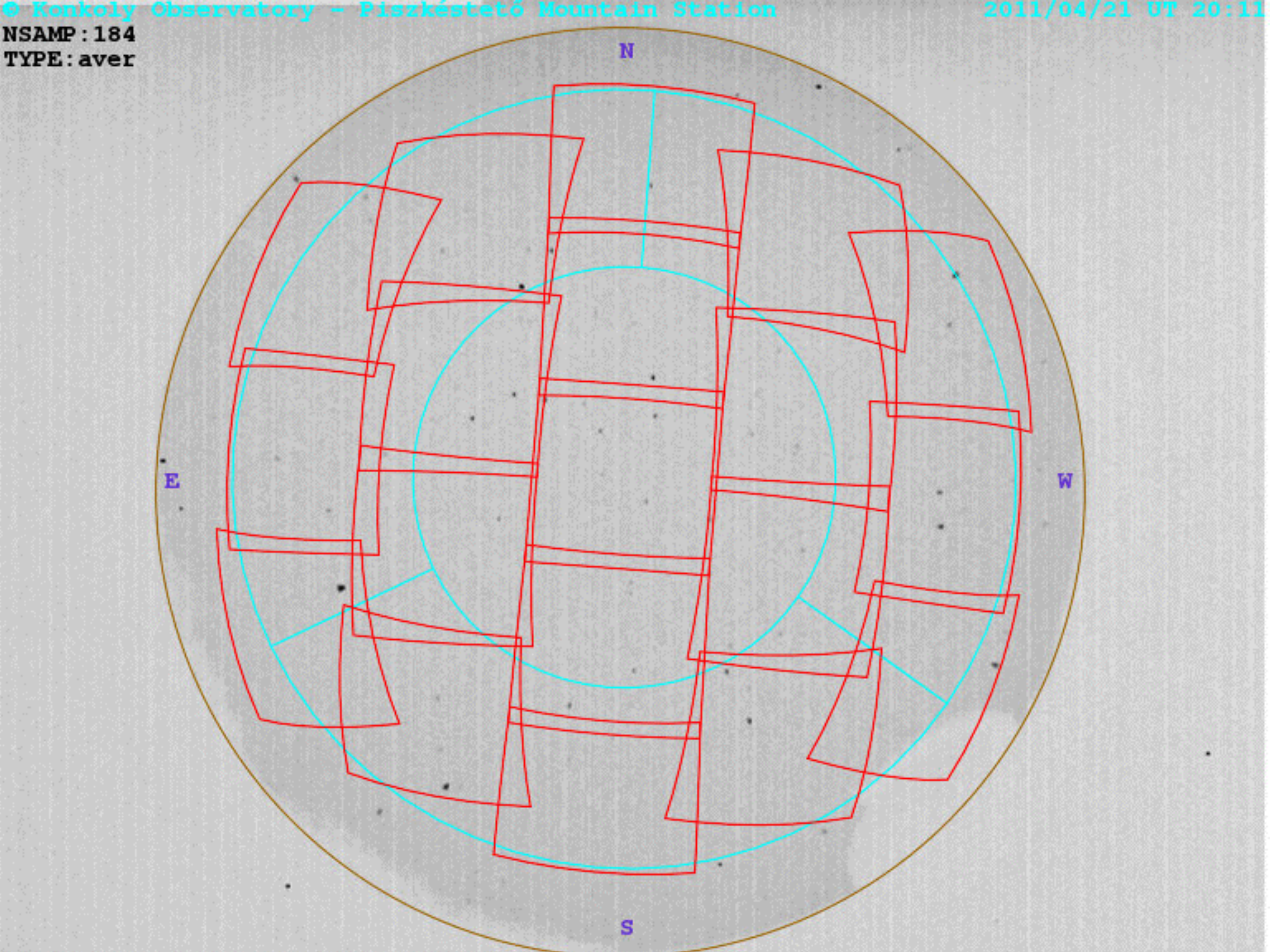}}
\end{center}
\caption{\emph{Left panel:} a simple visualization of the camera 
mount. The payload platform -- on which the 19 FLI Microline 
cameras have been mounted -- is shown to scale. Hence, 
the diameter of the platform is approximately 55\,cm
while the effective diameter including the cameras and lenses is
nearly 1\,m. The lower, fixed platform and the hexapod strut
drawings are merely figurative, but the expected distance between the
two platforms are roughly $25-30$\,cm (as it is implied by the figure scale).
The mosaic dome is partly shown also to scale,
as a transparent set of hexagonal elements. The size of the
hexagonal elements are roughly $23-25$\,cm.
 \emph{Right panel:}
the FOV of the 19 cameras shown on an (inverted) all-sky
image. Assuming a focal distance of $85$\,mm for the lenses and 
ML-16803 cameras as detectors, the FOV of each camera -- lens pair
will be roughly $26^{\circ}$. The placement and field orientation of 
the cameras are exactly the same as it is shown in the left panel.
The two concentric circles mark the $30^{\circ}$ and $60^{\circ}$ 
horizontal altitudes.}
\label{fig:flyseyedevice}
\end{figure*}

\section{Instrument design}
\label{sec:instrumentdesign}

The advance in consumer and computer electronics in recent years
allows us to build this {\it Fly's Eye} device from commercially available
and well-tested components with parameters that would not have been
possible even a few years ago. Hence, by exploiting these hardware
and optics, it is possible to design cost-effective instrumentation
for scientific purposes. In the following, we detail the properties
of specific cameras and lenses available in the marked from which
the proposed design can easily be built and deployed. 

The 19 cameras are mounted on a fixed assembly, i.e.,
the relative positions and field rotation angles are also kept fixed
throughout the observations. We intend to employ the very recent and
compact model ML-16803 of the FLI company and standard, commercially            
available Canon lenses with the focal length of $f=85\,{\rm mm}$ with a
rather fast focal ratio of $f/1.2$. This FOV will cover the sky 
all above the $h \ge 30^{\circ}$ horizontal altitude (i.e. half of the 
whole visible celestial sphere, up to airmasses $A\le 2$),
allowing the persistent survey of the sky with moderate imaging resolution.
According to the specifications of the $4{\rm k}\times 4{\rm k}$
KAF-16803 detectors 
that are used by the ML-16803 cameras, this setup yields a
resolution of $22^{\prime\prime}$/pixel. The resulting cumulative
optical light collecting power (i.e. the \'etendue) 
of the system will be nearly
$35-40$\,deg$^2$\,m$^2$, depending on the vignetting of the
lenses. This large value places the whole device among the group 
of the instruments with the highest optical attendance. 
The expected photometric precision expecting a cadence of 3 minutes 
is 4-500\,ppm for point sources of $r=10$\,magnitude while a precision
of 10\% is expected at $r=15$. The latter number can also be 
interpreted as a detection threshold with S/N=10. Due to its resolution
on the faint end, the photometric uncertainties are because of the
confusion of nearby sources. The astrometric and photometric
analysis of imaging data is performed
with the FITSH package\footnote{http://fitsh.szofi.net/}, 
found to be rather effective in wide-field
optical variability surveys (see \cite{pal2009phd,pal2012}).

A visualization of the 
mount design concept is displayed in Fig.~\ref{fig:flyseyedevice} 
along with the resulting sky coverage and the FOV of each 
camera-lens pair. 
The main concept of the camera design are a) to minimize the number
of moving parts and b) not to use specialized, uniquely designed and/or 
manufactured 
mechanical, optical or electronic components 
in the device. This second ``rule of thumb'' allows us 
to have spare parts of all of the necessary components that can be
replaced instantly upon a failure and hence, does not add a significant 
investment and maintenance cost. Therefore, one can expect a smooth and continuous operation 
of the camera system, since the simple design concepts allows a fast
replacement of broken parts. 

The camera platform will minimize the unique types of moving parts.
In the history of automated telescope surveys we clearly identified this
to be the main problem of reliable operation. 
The mount is planned to be based on a hexapod-design (also known as 
Steward-platform\footnote{See e.g. {http://en.wikipedia.org/wiki/Stewart\_platform}}),
that requires only identical mechanical elements and allows the desired 
motion  independently of the placement of the mount support base.
See also the left panel of Fig.~\ref{fig:flyseyedevice} of a schematic drawing 
of the hexapod mount. 

Sidereal tracking of the camera platform
is performed during the exposures while the local first equatorial
coordinates of the platform would exactly be the same throughout the
subsequent exposures. Therefore, during the image readout, the whole
platform is slewed back to its initial position and performs the same
apparent path in the next exposure and so on. 
The design for arranging the cameras
on the platform resembles the PASS instrument (\cite{deeg2004}).
In the PASS design, the cameras are fixed to the ground, hence there 
stellar images show trails: the essential improvement to PASS is the 
implementation of the sidereal tracking.

Since parameterization of the rotation via the pitch, roll and yaw axes 
does not imply any singularity (like a gimbal lock) for \emph{arbitrary}
small rotations (i.e. when the total rotation is $\rho \ll 90^{\circ}$),
this mount can be employed for sidereal tracking on arbitrary geographical
latitudes. Indeed, installing the mount to the poles of the Earth yields a pure
yaw rotation while installing the mount on the equator yields a pure
roll rotation (expecting the $x\pm$ axis pointing to north-south). 
On ``temperate'' latitudes, the sidereal rotation will
be a combination of yaw and roll, while the pitch rotation is required only
for correcting the polar alignment. The if the tracking accuracy should be
one tenth of the pixel resolution, 
$2^{\prime\prime}\approx10{\mu}{\rm rad}$, then the actuators must be
controlled with a precision of $\approx5{\mu}{\rm m}$ (expecting a
characteristic platform size and/or a retracted actuator length of $0.5$\,m).
This is feasible by our choice from the commercially available actuators.

\section{Further developments}
\label{sec:furtherdevelopment}

Since the {\it Fly's Eye} camera observes the sky above an altitude
of $30$ degrees, it simultaneously monitors almost exactly the one
fourth of the whole celestial sphere. From a temperate geographical 
latitude (e.g. from Hungary, at $\varphi\approx 47^\circ$), the Sun is
below the horizon more than $12^\circ$ within a time fraction of close
to $0.4$ on average throughout a year (but do not differ significantly
on other geographical locations). Hence, the system observes approximately
the $(1/4)\cdot0.4=10\%$ of the detectable events. A natural
way of increase this ratio is to install similar devices
at other locations on Earth. 
In addition, monitoring the sky simultaneously from distinct (and far)
locations significantly decreases the one-day aliases in the 
phase domains of periodic events. Moreover, synchronization in
image acquisition also aids the accurate data reduction since the 
overlapping regions observed by distinct devices have to be
exactly the same. This approach makes the characterization of systematic
noise sources much more easier. For instance, such a synchronization
can be accomplished by starting the exposures at every three minutes
in Greenwich sidereal time.

\begin{figure}
\begin{center}
\resizebox{!}{40mm}{\includegraphics{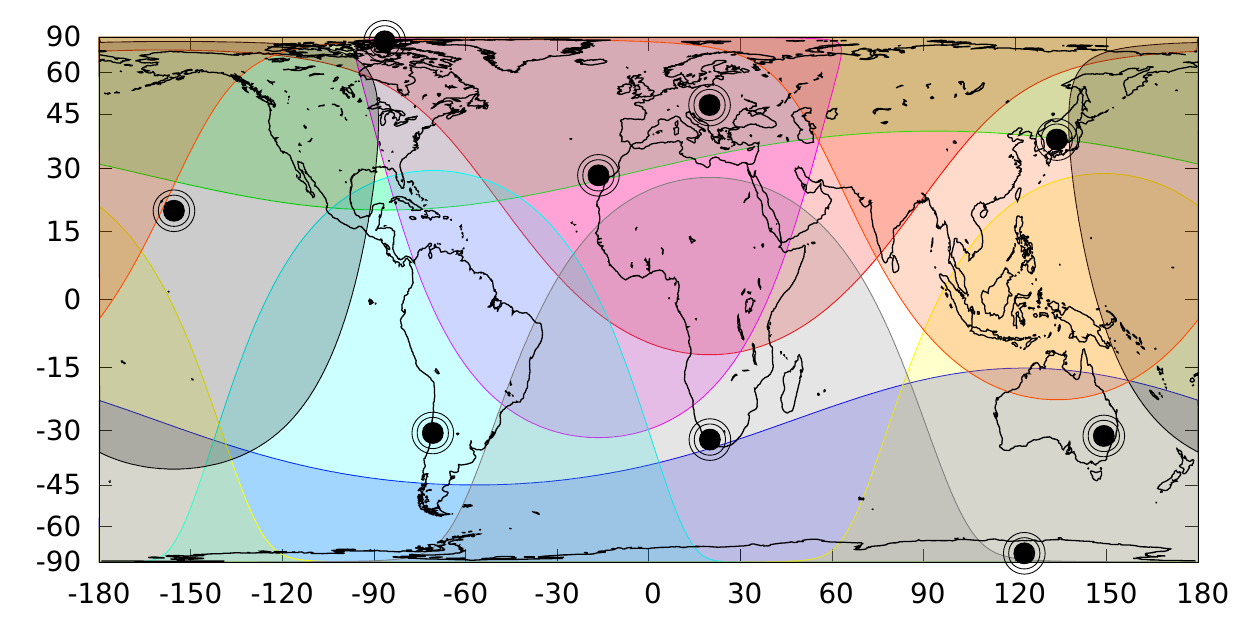}}
\end{center}
\caption{An example configuration and the yielded visibility coverage
of 9 {\it Fly's Eye} devices, distributed nearly homogeneously on the Earth
(using Lambert cylindrical equal-area projection). See text for further
details.}
\label{fig:coverage}
\end{figure}

Hence, we initiate further collaborations and seek
for other types of grants that will cover the costs of
building a network of these devices all around the world.
Negotiations have also been started with the staff of
Teide Observatory, Tenerife, Canary Islands. An example
configuration of $9$ devices located on various places on Earth
with well-known infrastructure suitable for installing 
astronomical instrumentation is displayed in Fig.~\ref{fig:coverage}.

\section{Scientific goals}
\label{sec:science}

The main goals of the proposed {\it Fly's Eye Network} project 
cover several topics in astrophysics. In the following, 
without attempting to be comprehensive, we list some of these sub-fields of
astrophysics. 

\subsection{Solar System} 

Even with its moderate resolution, 
the {\it Fly's Eye} device is capable to detect meteors and map 
these tracks with an effective resolution of $\sim10$\,m/pixel. Hence,
and due to the large light collecting power of the device, a more
accurate distribution of Solar System dust can be derived. In addition,
for the bright-end of the main-belt asteroid family members, an
unbiased sample will be available for their rotation and shape
properties (see also \cite{durech2011}).
These data are essential to understand the aspect
of Solar System dynamics and, more importantly, its evolution. 

Moreover,
nearby flybys of small bodies that are potentially hazardous to
the Earth can be traced (see e.g. the cases of \yu{} and \bx{}). 
Due to the continuous sampling, such information
is also recoverable in an \emph{a posteriori} manner, i.e. when deeper
surveys discover such an object and dynamical calculations confirm
a former approach in the FOV of one or more {\it Fly's Eye} device.

\subsection{Stellar and planetary systems}

Young stellar objects are complex astrophysical systems and show 
signs of both quasi-periodic and sudden 
transient, eruptive processes. By monitoring their intrinsic variability,
one is able to obtain several constraints regarding to the ongoing processes
(\cite{hartmann1996}, \cite{herbig2007}, \cite{abraham2009}, \cite{kospal2011}). 
Persistent monitoring of numerous young stellar 
objects or candidates for young stellar objects will reveal the nature of the 
currently unexplored domains of stellar birth. Since observing campaigns
are organized mostly on daily or yearly basis, the behaviour of
such systems is practically unknown on other timescales. 

Stars with magnetic activity show photometric variability on all the
time-domains of the planned instrument, from minutes through hours to
years, just like the Sun does (\cite{strassmeier2009}).
Continuous monitoring of the sky opens up a new research area for
active stars: the proposed device allows us to obtain good flare
statistics since flares occur on minutes-hours timescale,
and to monitor starspot evolution, differential rotation 
and activity cycles of the same star
(see e.g. \cite{hartman2011}, \cite{walkowicz2011}, \cite{olah2009}, 
\cite{vida2010}).

Observations of eclipsing binaries provide direct measurements of
stellar masses and radii that are essential to understand their evolution
and even the basic physical processes ongoing in the stellar cores
(\cite{latham2009}). Similarly to eclipsing binaries, transiting 
extrasolar planets are also expected to be discovered by the
{\it Fly's Eye Network}, since instruments with nearly similar types
of optics are found to be rather 
efficient (\cite{pollacco2004}, \cite{bakos2004}, \cite{pepper2007},
\cite{pal2008}, \cite{pal2009phd}).

\subsection{In the extragalactic environment}

Continuous monitoring of brighter supernovae in nearby galaxies
yield valuable data that can be exploited by combining
other kind of measurements. The {\it Fly's Eye} camera is capable to observe
the brightest supernovae directly even up to a month
during their peak brightness (see e.g. \cite{vinko2012}).
By combining images, it is possible to go even deeper in brightness
using more sophisticated ways of photometric techniques. 

\section{Summary}
\label{sec:summary}

This paper briefly summarized both the instrument design concepts
and the proposed series of scientific applications of an
all-sky monitoring device named {\it Fly's Eye Camera System}. Due to the
implementation of a hexapod-based camera platform, exactly the same instrument
can be installed independently from the current geographical location
as well as the whole setup does not need any kind of polar alignment. 
The robust design does not exploit any unique mechanical component
hence the operations are highly fault tolerant and the maintenance is easy. 
The resulting \'etendue of the optical setup in a {\it Fly's Eye} unit
is comparable to the largest available optical facilities, moreover,
a global network of such devices yields an \'etendue that is similar
to the current imaging system of the Large Synoptic Survey Telescope.

\acknowledgements
The ``Fly's Eye'' project is supported by the Hungarian Academy of
Sciences via the ``Lend\"ulet'' grant LP2012-31/2012. Additional support
is also received via the ESA grant PECS~98073. We thank 
Hans Deeg (PI of the PASS project) for the useful discussions. 
We also thank the careful comments of the anonymous referee. 

{}

\end{document}